\documentclass[aps,preprint]{revtex4}%
\usepackage{amsfonts}
\usepackage{amsmath}
\usepackage{amssymb}
\usepackage{graphicx}%
\begin{document}
\title{Dilatonic interpolation between Reissner-Nordstr\"{o}m and Bertotti-Robinson
spacetimes \ with physical consequences}
\author{S. Habib Mazharimousavi$^{\ast}$}
\author{M. Halilsoy$^{\dag}$}
\author{I. Sakalli$^{\sharp}$}
\author{O. Gurtug$^{\flat}$ }
\affiliation{Department of Physics, Eastern Mediterranean University, G. Magusa, north
Cyprus, Mersin-10, Turkey}
\affiliation{$^{\ast}$habib.mazhari@emu.edu.tr}
\affiliation{$^{\dagger}$mustafa.halilsoy@emu.edu.tr}
\affiliation{$^{\sharp}$izzet.sakalli@emu.edu.tr}
\affiliation{$^{\flat}$ozay.gurtug@emu.edu.tr}

\begin{abstract}
We give a general class of static, spherically symmetric, non-asymptotically
flat and asymptotically non-(anti) de Sitter black hole solutions in
Einstein-Maxwell-Dilaton (EMD) theory of gravity in 4-dimensions. In this
general study we couple a magnetic Maxwell field with a general dilaton
potential, while double Liouville-type potentials are coupled with the
gravity. We show that the dilatonic parameters play the key role in switching
between the Bertotti-Robinson and Reissner-Nordstr\"{o}m spacetimes. We study
the stability of such black holes under a linear radial perturbation, and in
this sense we find exceptional cases that the EMD black holes are unstable. In
continuation we give a detailed study of the spin-weighted harmonics in
dilatonic Hawking radiation spectrum and compare our results with the
previously known ones. Finally, we investigate the status of resulting naked
singularities of our general solution when probed with quantum test particles.

\end{abstract}
\maketitle

\section{Introduction}

We revisit the $4-$dimensional Einstein-Maxwell-Dilaton (EMD) theory and show
that there are still plenty of rooms available to contribute the subject.
Double Liouville potential and general dilaton coupling is considered to
obtain more general solutions with extra parameters and diagonal metric in the
theory. From the outset we remind that, depending on the relative parameters,
the double Liouville potential has the advantage of admitting local extrema
and critical points. The Higgs potential also shares such features, whereas
single Liouville potential lacks these properties. Double Liouville-type
potentials arise also when higher-dimensional theories are compactified to
$4-$dimensional spacetimes and expectedly bring in further richness. All known
solutions to date can be obtained \cite{GB,GHS,CHM} as particular limits of
our general solution, and it contains new solutions as well. In the most
general form our solution covers Reissner-Nordstrom (RN) type black holes and
Bertotti-Robinson (BR) spacetimes interpolated within the same metric.
Interpolation of two different solutions in general relativity is not a new
idea \cite{HALILSOY}. Particular limits of the dilatonic parameter yield the
RN and BR spacetimes. In between the two, the linear dilaton black hole (LDBH)
lies for the specific choice of the parameters. It is well-known that the near
horizon geometry of the extremal RN black hole yields the BR electromagnetic
universe. The latter \cite{BR} is important for various reasons: It is a
singularity free non-black hole solution which admits maximal symmetry and
finds application in conformal field theory correspondence (i.e. AdS/ CFT).
Particles in the BR universe move with uniform acceleration in a conformally
flat background. These features are mostly valid not only in $N=4$ but in
higher dimensions ($N>4$) as well. The topological structure of the BR
spacetime is still $AdS_{2}\times S^{N-2}$ in $N-$dimensions with the radius
of $S^{N-2}$ depending on the dimension of the space. Recently we have
extended the Maxwell part of the BR spacetime to cover the Yang-Mills (YM)
field and obtained common features that share with the Maxwell field
\cite{MH}. The dilatonic black hole solution involved in the general solution
obtained in this paper is non-asymptotically flat, therefore we expressed it
in terms of the quasi local mass ($M_{QL}$) \cite{BY}. The metric is regular
at horizons with only available singularity at $r=0$. Another feature is the
asymptotic $(r\rightarrow\infty)$ absence of (anti) de-Sitter property which
was discovered also within the context of different models \cite{CHM}. Our
general solution has been tested for stability against the radial, linear
perturbations. We found that presence of dilaton can trigger instability in
the RN black hole which is stable otherwise. Our analysis proves that the BR
sector remains manifestly stable against such perturbations. Thermodynamic
stability has also been discussed briefly by considering the specific heat of
the metric. Divergence in the specific heat for specific values of the
parameters signals phase transition in our thermodynamic system, i.e.,
topology change in the spacetime.

Next, we concentrate ourselves on the LDBH case and analyze the Hawking
temperature both from semi-classical and standard surface gravity methods
\cite{MSH,CFM}. We point out the contrasts between the two methods when there
are single and double horizons. The high frequency limit of the semi-classical
radiation spectrum method (\textit{SCRSM}) does not agree with the Hawking's
result. It is observed, as an interesting contribution in this work that the
coupling between scalar field charge and the magnetic charge of the spacetime
gives rise to spin-weighted spheroidal harmonics which plays a dominant role
in the difference. In the absence of such coupling, when the scalar field is
assumed chargeless for instance, similar analysis was carried out previously
and we had recovered the same results easily. It turns out that the very
existence of a spin-weighted spheroidal harmonics in the theory transforms a
divergent temperature spacetime to a finite one. We argue that such a behavior
may play a leading role in the detection of such LDBHs.

In the final section of the paper we appeal once more to the test scalar field
equation, but this time with the purpose to investigate the quantum nature of
the naked singularities. We identify first the particular solution that yields
horizonless naked singularity at $r=0$. By invoking the Horowitz-Marolf
\cite{HOROWITZ} criterion on quantum nature of classical singularities we
explore under which set of parameters classically singular but quantum
mechanically regular metrics can occur in our general solution.

The organization of our paper is as follows. In Sec. II we introduce our
action, field equations and obtain the general solution. Sec. III singles out
the linear dilaton case and investigates the stability of our general
solution. Application of the \textit{SCRSM} and its connection with the
Hawking temperature is employed in Sec. IV. Sec. V discusses the status of
naked singularities from quantum picture. We summarize our results in
conclusion which appears in Sec. VI.

\section{Field Equations and the metric ansatz for EMD gravity}

The $4-$dimensional action in the EMD theory is given by $(8\pi G=1)$%
\begin{equation}
S=\int d^{4}x\sqrt{-g}\left(  \frac{1}{2}R-\frac{1}{2}\partial_{\mu}%
\phi\partial^{\mu}\phi-V\left(  \phi\right)  -\frac{1}{2}W\left(  \phi\right)
(F_{\lambda\sigma}F^{\lambda\sigma})\right)  ,\text{\ }%
\end{equation}
where%
\begin{equation}
V\left(  \phi\right)  =V_{1}e^{\beta_{1}\phi}+V_{2}e^{\beta_{2}\phi},\text{
\ \ }W\left(  \phi\right)  =\lambda_{1}e^{-2\gamma_{1}\phi}+\lambda
_{2}e^{-2\gamma_{2}\phi}.
\end{equation}
$\phi$ refers to the dilaton scalar potential and $\gamma_{i}$ denotes the
dilaton parameter, $\lambda_{i}$ is a constant and $V\left(  \phi\right)  $ is
a double Liouville-type potential. We note that we exclude the simultaneous
values $\beta_{1}=\beta_{2}$ and $\gamma_{1}=\gamma_{2}$ in general, since
these particular values lead to the already known cases.

Let us remark that although double Liouville potential in $V\left(
\phi\right)  ,$ which renders local minima, necessary for construction of
vacuum states possible, the similar choice for $W\left(  \phi\right)  $ seems
less appealing. It will be justified from the exact solutions below, however,
that there are asymptotics which remains inaccessible by the choice of a
single Liouville term in $W\left(  \phi\right)  .$ Stated otherwise, at both
asymptotes of $r=0$ and $r=\infty$ (or $\tilde{r}=0$ and $\tilde{r}=\infty$
for LDBH) dilatonic coupling to the magnetic field becomes much stronger.
Choosing a single Liouville potential simply looses the strength at one end of
the range. Besides, it is all a matter of choice to set $\lambda_{1}\left(
\lambda_{2}\right)  =0,$ which makes the dilatonic coupling asymptotically
free. In the LDBH case as it will be proved, if we set $\lambda_{1}=0,$ we
shall remove the possibility of an inner (Cauchy) horizon which justifies the
advantages and motivation for choosing the double Liouville-type potential in
$W\left(  \phi\right)  .$ In (1) $R$ is the usual Ricci scalar and
$\mathbf{F}=\frac{1}{2}F_{\mu\nu}dx^{\mu}\wedge dx^{\nu}$ is the Maxwell
$2-$form (with $\wedge$ indicating the wedge product) given by
\begin{equation}
\mathbf{F}=\mathbf{dA,}%
\end{equation}
for $\mathbf{A}=A_{\mu}dx^{\mu}$, the potential $1-$form. Our pure magnetic
potential with charge $Q$, which is given by%
\begin{equation}
\mathbf{A}=-Q\cos\theta\ d\varphi,\text{ }%
\end{equation}
leads to%
\begin{equation}
\mathbf{F}=Q\sin\theta\ d\theta\wedge d\varphi.
\end{equation}
Let us note that with the present choice of $W\left(  \phi\right)  $ the
electric-magnetic symmetry that exists in the standard dilatonic coupling,
i.e., $\lambda_{1}\left(  \lambda_{2}\right)  =0,$ is no more valid. Our
choice in this paper relies entirely on the magnetic choice. Variations of the
action with respect to the gravitational field $g_{\mu\nu}$ and the scalar
field $\phi$ lead, respectively to the EMD field equations
\begin{gather}
R_{\mu\nu}=\partial_{\mu}\phi\partial_{\nu}\phi+V\left(  \phi\right)
g_{\mu\nu}+W\left(  \phi\right)  \left(  2F_{\mu\lambda}F_{\nu}^{\ \lambda
}-\frac{1}{2}F_{\lambda\sigma}F^{\lambda\sigma}g_{\mu\nu}\right)  ,\\
\nabla^{2}\phi-V^{\prime}\left(  \phi\right)  -\frac{1}{2}W^{\prime}\left(
\phi\right)  (F_{\lambda\sigma}F^{\lambda\sigma})=0,\\
\left(  \prime\equiv\frac{d}{d\phi}\right)  ,\nonumber
\end{gather}
where $R_{\mu\nu}$ is the Ricci tensor. Variation with respect to the gauge
potential $\mathbf{A}$ yields the Maxwell equation%

\begin{equation}
\mathbf{d}\left(  W\left(  \phi\right)  ^{\star}\mathbf{F}\right)  =0,
\end{equation}
in which the hodge star $^{\star}$ means duality.

\subsection{Ansatz and the Solutions:}

Our ansatz line element for EMD gravity is chosen to be
\begin{equation}
ds^{2}=-f\left(  r\right)  dt^{2}+\frac{1}{f\left(  r\right)  }dr^{2}%
+R(r)^{2}\left(  d\theta^{2}+\sin^{2}\theta d\varphi^{2}\right)  ,
\end{equation}
with $f\left(  r\right)  $ and $R\left(  r\right)  $ only function of $r$
while the Maxwell invariant takes the form%
\begin{equation}
F_{\lambda\sigma}F^{\lambda\sigma}=\frac{2Q^{2}}{R^{4}}.
\end{equation}
The Maxwell equation (8) is satisfied automatically and the field equations
become%
\begin{align}
\nabla^{2}\phi &  :=\frac{1}{R^{2}}\left(  R^{2}f\phi^{\prime}\right)
^{\prime}=V^{\prime}\left(  \phi\right)  +\frac{1}{2}W^{\prime}\left(
\phi\right)  (F_{\lambda\sigma}F^{\lambda\sigma}),\\
R_{t}^{t}  &  :=-\frac{\left(  f^{\prime}R^{2}\right)  ^{\prime}}{2R^{2}%
}=V\left(  \phi\right)  -\frac{W\left(  \phi\right)  }{2}(F_{\lambda\sigma
}F^{\lambda\sigma}),\\
R_{r}^{r}  &  :=-\frac{2fR^{\prime\prime}}{R}-\frac{\left(  f^{\prime}%
R^{2}\right)  ^{\prime}}{2R^{2}}=f\phi^{\prime2}+V\left(  \phi\right)
-\frac{W\left(  \phi\right)  }{2}(F_{\lambda\sigma}F^{\lambda\sigma}),\\
R_{\theta}^{\theta}  &  =R_{\varphi}^{\varphi}:=\frac{1-\left(  fRR^{\prime
}\right)  ^{\prime}}{R^{2}}=V\left(  \phi\right)  +\frac{2Q^{2}W\left(
\phi\right)  }{R\left(  r\right)  ^{4}}-\frac{W\left(  \phi\right)  }%
{2}(F_{\lambda\sigma}F^{\lambda\sigma}),
\end{align}
in which a prime stands for derivative with respect to the argument of the
function. We start with an ansatz for $R(r)$ as%
\begin{equation}
R\left(  r\right)  =Ae^{\eta\phi}%
\end{equation}
in which $A$ and $\eta$ are constants to be found. Substitution in (12) and
(13), implies
\begin{equation}
\phi\left(  r\right)  =\frac{2\eta}{2\eta^{2}+1}\ln r.
\end{equation}
Finally by putting these results into Eq.s (11) and (14) one finds that by
setting
\begin{equation}
\eta=-\frac{1}{\alpha\sqrt{2}},
\end{equation}
and
\begin{gather}
\gamma_{1}=-\frac{\alpha}{\sqrt{2}},\text{ \ \ }\gamma_{2}=\frac{1}%
{\alpha\sqrt{2}},\\
\beta_{1}=\sqrt{2}\alpha,\text{ \ \ }\beta_{2}=\frac{\sqrt{2}}{\alpha
},\nonumber
\end{gather}
a general solution for $f(r)$ reads
\begin{align}
f\left(  r\right)   &  =\left(  1+\alpha^{2}\right)  ^{2}\left[  \frac
{Q^{2}\lambda_{1}r^{\frac{-2}{1+\alpha^{2}}}}{\left(  1+\alpha^{2}\right)
A^{4}}\right.  +\\
&  \left.  \left(  \frac{Q^{2}\lambda_{2}}{A^{4}}-V_{2}\right)  \frac
{r^{\frac{2\alpha^{2}}{1+\alpha^{2}}}}{\left(  1+\alpha^{2}\right)  \alpha
^{2}}-\frac{V_{1}r^{\frac{2}{1+\alpha^{2}}}}{3-\alpha^{2}}-Mr^{-\frac
{1-\alpha^{2}}{1+\alpha^{2}}}\right]  ,\nonumber
\end{align}
with the constraint condition%
\begin{equation}
-V_{2}\left(  1-\alpha^{2}\right)  A^{4}-\alpha^{2}A^{2}+\lambda_{2}%
Q^{2}\left(  1+\alpha^{2}\right)  =0.
\end{equation}
Herein $M$ is a mass-related integration constant and $\alpha$ and $A$ are
constants that will serve to parametrize the solution. We note that in case
that we are interested in the Newtonian limit, when $\alpha=0=\lambda
_{2}=V_{2},$ and $r\rightarrow\infty,$ we must choose $M\rightarrow2M,$ so
that $M$ represents the Newtonian mass. Therefore the dilatonic function
$\phi$, Liouville potential $V$ and $W$ in terms of $\alpha$ become
\begin{gather}
\phi\left(  r\right)  =-\frac{\alpha\sqrt{2}}{1+\alpha^{2}}\ln r,\text{
\ \ }R\left(  r\right)  =Ar^{\frac{1}{1+\alpha^{2}}},\nonumber\\
V=V_{1}r^{\frac{-2\alpha^{2}}{1+\alpha^{2}}}+V_{2}r^{\frac{-2}{1+\alpha^{2}}%
},\text{ }W=\lambda_{1}r^{\frac{-2\alpha^{2}}{1+\alpha^{2}}}+\lambda
_{2}r^{\frac{2}{1+\alpha^{2}}}.\text{\ \ }\nonumber
\end{gather}
We remark that the solution (16-20) is the general diagonal solution that
covers all particular solutions of this kind known so far. For arbitrary value
of $\alpha,$ other then $0,$ $1$ and $\infty,$ it yields a new solution in
accordance with our ansatz. It is observed also that our metric and potentials
are invariant under $\alpha\rightarrow-\alpha,$ whereas $\phi\rightarrow
-\phi.$ The asymptotic behavior of the metric function, $f\left(  r\right)  $
and other limiting cases can be summarized as follows%
\begin{equation}
\lim_{r\rightarrow\infty}f\left(  r\right)  \rightarrow\left\{
\begin{tabular}
[c]{ll}%
$\left(  1+\alpha^{2}\right)  ^{2}\left(  -\frac{V_{1}r^{\frac{2}{1+\alpha
^{2}}}}{3-\alpha^{2}}\right)  $ & $0\leq\alpha^{2}<1$\\
$2\left(  \frac{Q^{2}\lambda_{2}}{A^{4}}-V_{1}-V_{2}\right)  r$ & $\alpha
^{2}=1$\\
$\left(  \frac{1+\alpha^{2}}{\alpha^{2}}\right)  \left(  \frac{Q^{2}%
\lambda_{2}}{A^{4}}-V_{2}\right)  r^{\frac{2\alpha^{2}}{1+\alpha^{2}}}$ &
$1<\alpha^{2}$%
\end{tabular}
\ \ \ \ \ \ \ \right.
\end{equation}%
\begin{equation}
\lim_{r\rightarrow0^{+}}f\left(  r\right)  \rightarrow\left(  1+\alpha
^{2}\right)  \left(  \frac{Q^{2}\lambda_{1}}{A^{4}r^{\frac{2}{1+\alpha^{2}}}%
}\right)  .
\end{equation}
The case $\alpha^{2}=1$ will be studied separately, while the case $\alpha
^{2}=0,$ with the choice of $\lambda_{2}=0,$ $V_{2}=0,$ and $A=1=\lambda_{1}$
leads to%
\begin{align}
f\left(  r\right)   &  =1-\frac{V_{1}}{3}r^{2}-\frac{M}{r}+\frac{Q^{2}}{r^{2}%
},\nonumber\\
R\left(  r\right)   &  =r,\phi=0,
\end{align}
which corresponds to the action%
\begin{equation}
S_{\alpha^{2}=0}=\int d^{4}x\sqrt{-g}\left(  \frac{1}{2}R-V_{1}-\frac{1}%
{2}(F_{\lambda\sigma}F^{\lambda\sigma})\right)  .
\end{equation}
This is recognized as the $4-$dimensional action in the EM theory with the
solution representing a RN black hole with a cosmological constant. Another
limiting case of interest consists of the case with $\alpha^{2}\rightarrow
\infty,$ $\lambda_{2}=1,$ with the action%
\begin{equation}
S_{\alpha^{2}=\infty}=\int d^{4}x\sqrt{-g}\left(  \frac{1}{2}R-V_{2}-\frac
{1}{2}(F_{\lambda\sigma}F^{\lambda\sigma})\right)  ,
\end{equation}
leading to the solution%
\begin{gather}
f\left(  r\right)  =\left(  \frac{Q^{2}}{A^{4}}-V_{2}\right)  r^{2}-\tilde
{M}r,\\
A^{2}\left(  V_{2}A^{2}+1\right)  =Q^{2},\nonumber\\
R\left(  r\right)  =A,\text{ \ \ }\phi\left(  r\right)  =0,\nonumber
\end{gather}
in which $\tilde{M}$ is the mass related integration constant. Here also we
have a $4-$dimensional action in the EM theory with cosmological constant but
the metric function represents a BR space time.

By looking at the asymptotic behaviors of the general solution one finds that
$0\leq\alpha^{2}<1$ and $1<\alpha^{2}$ correspond to the cases of RN \ and BR
solutions, respectively. Here $\alpha^{2}=1$ acts much like a phase transition
which changes the structure of space time from RN into BR. \ The thermodynamic
instability from the expression of specific heat capacity $C_{Q},$ (Eq. (58)
given below) justifies this fact. It is quite interesting to see what will be
the answer if one chooses $\alpha^{2}=1.$ In the next section we concentrate
on this critical value for $\alpha^{2}.$

\section{The Linear Dilaton}

From the asymptotic behavior of the metric function one may see that
$\alpha^{2}=1$ is a critical value and the behavior of spacetime changes. In
this chapter we only concentrate on this specific value for $\alpha^{2}$, and
will be referred to as linear dilaton. The general solution after this setting
reads%
\begin{gather}
-\gamma_{1}=\gamma_{2}=\frac{1}{\sqrt{2}},\text{ \ }\\
\beta_{1}=\beta_{2}=\sqrt{2},\text{ \ }\nonumber\\
\phi\left(  r\right)  =-\frac{1}{\sqrt{2}}\ln r,\text{ \ \ }R\left(  r\right)
=A\sqrt{r},\nonumber\\
V=\frac{\tilde{V}}{r},\text{ \ }W=\frac{\lambda_{1}}{r}+\lambda_{2}%
r,\nonumber\\
A^{2}=2\lambda_{2}Q^{2},\text{ \ \ }\left(  \lambda_{2}>0\right) \nonumber
\end{gather}%
\begin{equation}
f\left(  r\right)  =\left[  \frac{\lambda_{1}}{\lambda_{2}A^{2}r}+\left(
\frac{1}{A^{2}}-2\tilde{V}\right)  r-\tilde{M}\right]  ,
\end{equation}
where $\tilde{V}=V_{1}+V_{2}.$

In order to explore the physical properties of the linear dilaton case we
perform the transformation $R\left(  r\right)  =A\sqrt{r}\rightarrow
\widetilde{r}.$ This transforms the metric into,%

\begin{equation}
ds^{2}=-f(\widetilde{r})dt^{2}+\frac{4\widetilde{r}^{2}}{A^{4}f(\widetilde
{r})}d\widetilde{r}^{2}+\widetilde{r}^{2}d\Omega^{2},
\end{equation}
in which%

\begin{equation}
f(\widetilde{r})=\frac{1}{\widetilde{r}^{2}}\left(  \left(  \frac{1}{A^{2}%
}-2\tilde{V}\right)  \frac{\widetilde{r}^{4}}{A^{2}}-M_{QL}\widetilde{r}%
^{2}+\frac{\lambda_{1}}{\lambda_{2}}\right)  ,
\end{equation}
where the mass $M_{QL}$ denotes the quasilocal mass whose general definition
is given below in Eq. (46). Other related parameters transform into the
following forms,%

\begin{align}
\phi\left(  \widetilde{r}\right)   &  =\sqrt{2}\ln\left(  \frac{A}%
{\widetilde{r}}\right)  ,\\
V(\widetilde{r})  &  =\frac{A^{2}}{\widetilde{r}^{2}}\left(  V_{1}%
+V_{2}\right)  ,\text{ \ \ \ \ \ \ \ \ }W\left(  \widetilde{r}\right)
=\frac{\lambda_{1}A^{2}}{\widetilde{r}^{2}}+\frac{\lambda_{2}\widetilde{r}%
^{2}}{A^{2}}.\text{ }\nonumber
\end{align}
The location of horizons can be found if we set the metric function
$g_{tt}=0.$ The solution is%

\begin{equation}
\widetilde{r}_{h}=\frac{1}{\sqrt{2a}}\sqrt{M_{QL}\pm\sqrt{M_{QL}^{2}-4ac}},
\end{equation}
where%

\begin{equation}
a=\left(  \frac{1}{A^{2}}-2\tilde{V}\right)  \frac{1}{A^{2}},\text{
\ \ \ }\ c=\frac{\lambda_{1}}{\lambda_{2}}.
\end{equation}

The linear dilaton solution admits single or double-horizons if the parameters
are chosen appropriately. Another possible case is the extremal limit that
occurs if $M_{QL}^{2}=4ac.$ The horizon in this particular case is given by
$\widetilde{r}_{h}=\sqrt{\frac{M_{QL}}{2a}}.$ The double horizon case occurs
if the parameters simultaneously satisfy $M_{QL}>\sqrt{M_{QL}^{2}-4ac}$ and
$M_{QL}^{2}>4ac.$ This choice leads to the horizons%

\begin{align}
\widetilde{r}_{+}  &  =\sqrt{\frac{M_{QL}+\sqrt{M_{QL}^{2}-4ac}}{2a}},\\
\widetilde{r}_{-}  &  =\sqrt{\frac{M_{QL}-\sqrt{M_{QL}^{2}-4ac}}{2a}%
}.\nonumber
\end{align}

The relations between the parameters and double-Liouville-type potentials in
the formation of black holes becomes evident if one looks for the critical
case. This is the case when $M_{QL}=\sqrt{M_{QL}^{2}-4ac},$ which follows that
$2\tilde{V}=\frac{1}{A^{2}}.$ Hence if $2\tilde{V}<\frac{1}{A^{2}},$ no
horizon forms and the central singularity $\widetilde{r}=0$ becomes a
\textit{naked} singularity.\ It can easily be seen that for $\lambda_{1}=0$,
or for the single Liouville-type potential in $W\left(  \phi\right)  ,$ we
have automatically single, outer event horizon alone. Another interesting
property is in the behavior of the curvature scalar $R$. The curvature scalar
for the metric function (29) is,%

\begin{equation}
R=-\frac{4\widetilde{r}^{4}\left(  aA^{4}-1\right)  -A^{4}\left(
a\widetilde{r}^{4}-M_{QL}\widetilde{r}^{2}+c\right)  }{2\widetilde{r}^{6}}.
\end{equation}
Note that the curvature scalar is finite at the location of horizons.
Furthermore, when $\widetilde{r}\rightarrow\infty$ , the Kretschmann and
curvature scalars, the Liouville-type potentials and the coupling term of
dilaton with Maxwell field all vanish. The mass and charge are finite and the
dominant field is gravity with finite curvature. Consequently, the solution
given in Eq. (29) is well-behaved. However, the $Q=0$ limit does not exist.

\subsection{Linear stability analysis of the general solution}

By employing a similar method used by Yazadjiev \cite{YAZ} we investigate the
stability of the possible EMD solution, in terms of a linear, radial
perturbation. To do so we assume that our dilatonic scalar field $\phi\left(
r\right)  $ changes into $\phi_{\circ}\left(  r\right)  +\psi\left(
t,r\right)  ,$ in which $\psi\left(  t,r\right)  $ is very weak compared to
the original dilaton field $\phi_{\circ}\left(  r\right)  $ and we call it the
perturbed term. As a result we choose our perturbed metric as
\begin{equation}
ds^{2}=-f\left(  r\right)  e^{\Gamma\left(  t,r\right)  }dt^{2}+e^{\chi\left(
t,r\right)  }\frac{dr^{2}}{f\left(  r\right)  }+R\left(  r\right)  ^{2}%
d\Omega_{2}^{2}.
\end{equation}
One should notice that, since our gauge potentials are magnetic, the Maxwell
equations (Eq.(8)) are satisfied. The linearized version of the field
equations (11-14) plus one extra term for $R_{tr}$ are given by%
\begin{gather}
R_{tr}:\frac{\chi_{t}\left(  t,r\right)  R^{\prime}\left(  r\right)
}{R\left(  r\right)  }=\partial_{r}\phi_{\circ}\left(  r\right)  \partial
_{t}\psi\left(  t,r\right)  \\
\nabla_{\circ}^{2}\psi-\chi\nabla_{\circ}^{2}\phi_{\circ}+\frac{1}{2}\left(
\Gamma-\chi\right)  _{r}\phi_{\circ}^{\prime}f-\partial_{\phi_{\circ}}%
^{2}V\left(  \phi_{\circ}\right)  \psi=\frac{Q^{2}}{R\left(  r\right)  ^{4}%
}\partial_{\phi_{\circ}}^{2}W\left(  \phi_{\circ}\right)  \psi\\
R_{\theta\theta}:\left(  1-R_{\circ\theta\theta}\right)  \chi-\frac{1}%
{2}RR^{\prime}f\left(  \Gamma-\chi\right)  _{r}=\left(  R^{2}\partial
_{\phi_{\circ}}V\left(  \phi_{\circ}\right)  +\frac{Q^{2}}{R^{2}}%
\partial_{\phi_{\circ}}W\left(  \phi_{\circ}\right)  \right)  \psi
\end{gather}
in which a lower index $_{\circ}$ represents the quantity in the unperturbed
metric. First equation in this set implies%
\begin{equation}
\chi\left(  t,r\right)  =\frac{1}{\eta}\psi\left(  t,r\right)
\end{equation}
which after making substitutions in the two latter equations and eliminating
the $\left(  \Gamma-\chi\right)  _{r}$ one finds%
\begin{equation}
\nabla_{\circ}^{2}\psi\left(  t,r\right)  -U\left(  r\right)  \psi\left(
t,r\right)  =0
\end{equation}
where%
\begin{equation}
U\left(  r\right)  =\frac{2}{r^{\frac{2}{1+\alpha^{2}}}}\left\{  \frac
{\alpha^{2}}{A^{2}}+\left(  \frac{Q^{2}\lambda_{2}}{A^{4}}+V_{2}\right)
\left(  \frac{1-\alpha^{4}}{\alpha^{2}}\right)  \right\}  .
\end{equation}
To get these results we have implicitly used the constraint (20) on $A$. Again
by imposing the same constraint , one can show that $U\left(  r\right)  $ is
positive. It is not difficult to apply the separation method on (41) to get
\begin{equation}
\psi\left(  t,r\right)  =e^{\pm\epsilon t}\zeta\left(  r\right)  ,\text{
\ \ }\nabla_{\circ}^{2}\zeta\left(  r\right)  -U_{eff}\left(  r\right)
\zeta\left(  r\right)  =0,\text{ \ \ }U_{eff}\left(  r\right)  =\left(
\frac{\epsilon^{2}}{f}+U\left(  r\right)  \right)  ,
\end{equation}
where $\epsilon$ is a constant. If one shows that the effective potential
$U_{eff}\left(  r\right)  $ is positive for any real value for $\epsilon$ it
means that there exists a solution for $\zeta\left(  r\right)  $ which is not
bounded. In other words by the linear perturbation our black hole solution is
stable for any value of $\epsilon.$

But in our case one must be careful. For instance let's go back to the general
solution (19) and set $V_{1}=0,$
\begin{equation}
f\left(  r\right)  =\frac{\left(  1+\alpha^{2}\right)  }{r^{\frac{2}%
{1+\alpha^{2}}}}\left\{  \left(  \frac{Q^{2}\lambda_{2}}{A^{4}}-V_{2}\right)
\frac{r^{2}}{\alpha^{2}}-M\left(  1+\alpha^{2}\right)  r+\frac{Q^{2}%
\lambda_{1}}{A^{4}}\right\}  ,
\end{equation}
this solution may have double horizons, single horizon (extremal) or no
horizon. These depend on the values of the parameters. One may notice that
this solution is a non-asymptotically flat metric and therefore the ADM mass
is not defined in general. Following the quasilocal mass formalism introduced
by Brown and York \cite{BY} it is known that, a spherically symmetric
$N-$dimensional metric solution as
\begin{equation}
ds^{2}=-F\left(  R\right)  ^{2}dt^{2}+\frac{dR^{2}}{G\left(  R\right)  ^{2}%
}+R^{2}d\Omega_{N-2}^{2},
\end{equation}
admits a quasilocal mass $M_{QL}$ defined by \cite{MH, BY}%
\begin{equation}
M_{QL}=\frac{N-2}{2}R_{B}^{N-3}F\left(  R_{B}\right)  \left(  G_{ref}\left(
R_{B}\right)  -G\left(  R_{B}\right)  \right)  .
\end{equation}
Here $G_{ref}\left(  R\right)  $ is an arbitrary non-negative reference
function, which yields the zero of the energy for the background spacetime,
and $R_{B}$ is the radius of the spacelike hypersurface boundary. Applying
this formalism to the solution (44), one obtains the horizon $M$ in terms of
$M_{QL}$ as%
\begin{equation}
M=\frac{2}{\left(  1+\alpha^{2}\right)  A^{2}}M_{QL},
\end{equation}
after which the metric function becomes%
\begin{equation}
f\left(  r\right)  =\frac{\left(  1+\alpha^{2}\right)  }{r^{\frac{2}%
{1+\alpha^{2}}}}\left\{  \left(  \frac{Q^{2}\lambda_{2}}{A^{4}}-V_{2}\right)
\frac{r^{2}}{\alpha^{2}}-\frac{2M_{QL}}{A^{2}}r+\frac{Q^{2}\lambda_{1}}{A^{4}%
}\right\}  .
\end{equation}
Indeed, since we wish to cover all known solutions in the literature of this
kind, we consider $\lambda_{i},$ $M_{QL}\geq0,$ and
\begin{equation}
\left(  \frac{Q^{2}\lambda_{2}}{A^{4}}-V_{2}\right)  \geq0.
\end{equation}
This condition together with Eq. (20) give a transparent view of
$U_{eff}\left(  r\right)  .$ In other words, after simplification, one can
rewrite $U\left(  r\right)  $ as
\begin{equation}
U\left(  r\right)  =\frac{2}{r^{\frac{2}{1+\alpha^{2}}}}\left\{  \left(
\frac{Q^{2}\lambda_{2}}{A^{4}}-V_{2}\right)  +\left(  \frac{Q^{2}\lambda_{2}%
}{A^{4}}+V_{2}\right)  \frac{1}{\alpha^{2}}\right\}  ,
\end{equation}
which reveals for $-\frac{Q^{2}\lambda_{2}}{A^{4}}\leq V_{2}\leq\frac
{Q^{2}\lambda_{2}}{A^{4}},$ $U\left(  r\right)  $ and then $U_{eff}\left(
r\right)  $ are positive, which means \ that the corresponding metric is
stable. But for $V_{2}<-\frac{Q^{2}\lambda_{2}}{A^{4}},$ if $\alpha^{2}%
<\alpha_{critical}^{2}$ where%
\begin{equation}
\alpha_{critical}^{2}=\frac{\left\vert V_{2}\right\vert -\frac{Q^{2}%
\lambda_{2}}{A^{4}}}{\left\vert V_{2}\right\vert +\frac{Q^{2}\lambda_{2}%
}{A^{4}}},
\end{equation}
then $U\left(  r\right)  $ gets negative value and therefore our solution
faces an instability condition. Here it is interesting to note that
$\alpha_{critical}^{2}<1$ belongs to the RN type black hole solutions, i.e. BR
type solution is automatically stable for any value of $\alpha^{2}.$

The general solution reveals another interesting case after we set $V_{1}=0,$
and $\lambda_{2}=0$ i.e.
\begin{equation}
f\left(  r\right)  =\frac{\left(  1+\alpha^{2}\right)  }{r^{\frac{2}%
{1+\alpha^{2}}}}\left\{  -\frac{V_{2}}{\alpha^{2}}r^{2}-\frac{2M_{QL}}{A^{2}%
}r+\frac{Q^{2}\lambda_{1}}{A^{4}}\right\}  .
\end{equation}
Upon choosing $V_{2}<0$ $\left(  V_{2}>0\right)  $ this admits the effective
potential%
\begin{equation}
U\left(  r\right)  =\frac{2\left\vert V_{2}\right\vert }{r^{\frac{2}%
{1+\alpha^{2}}}}\left(  \frac{\alpha^{2}-1}{\alpha^{2}}\right)
\end{equation}
which clearly from (20), for $\alpha^{2}<1\left(  \alpha^{2}>1\right)  $
manifests an unstable black hole solution. As a result we observe that a
stable RN black hole becomes unstable under certain conditions in the presence
of a dilaton and a Liouville potential.

\subsection{Thermodynamic stability}

Concerning the solution (19), we set the parameters $\lambda_{1}=\lambda
_{2}=1$ and $V_{1}=0$ to get
\begin{equation}
f\left(  r\right)  =\frac{\left(  1+\alpha^{2}\right)  }{r^{\frac{2}%
{1+\alpha^{2}}}}\left\{  \left(  \frac{Q^{2}}{A^{4}}-V_{2}\right)  \frac
{r^{2}}{\alpha^{2}}-\frac{2M_{QL}}{A^{2}}r+\frac{Q^{2}}{A^{4}}\right\}
\end{equation}
which in terms of the radius of horizon $r_{h}$ one finds the quasilocal mass
as
\begin{equation}
M_{QL}=\frac{r_{h}^{2}\left(  Q^{2}-V_{2}A^{4}\right)  +Q^{2}\alpha^{2}%
}{2A^{2}\alpha^{2}r_{h}}.
\end{equation}
The Hawking temperature
\begin{equation}
T_{H}=\frac{f^{\prime}\left(  r_{h}\right)  }{4\pi}=\frac{\left(  1+\alpha
^{2}\right)  \left[  r_{h}^{2}\left(  A^{2}-2Q^{2}\right)  -Q^{2}\left(
1-\alpha^{2}\right)  \right]  }{4\left(  1-\alpha^{2}\right)  A^{4}\pi
r_{h}^{\frac{3+\alpha^{2}}{1+\alpha^{2}}}}.
\end{equation}
and the Bekenstein-Hawking entropy
\begin{equation}
S=\frac{\mathfrak{a}}{4}=\pi r_{h}^{2},
\end{equation}
where $\mathfrak{a}$ is the area of the black hole, together lead to the heat
capacity $C_{Q}$ for constant $Q$ as%
\begin{equation}
C_{Q}=T_{H}\left(  \frac{\partial S}{\partial T_{H}}\right)  _{Q}%
=\frac{\left(  \alpha^{2}+1\right)  \left[  r_{h}^{2}\left(  2-\left(
\frac{A}{Q}\right)  ^{2}\right)  +1-\alpha^{2}\right]  }{\left(  \alpha
^{2}-1\right)  \left[  r_{h}^{2}\left(  2-\left(  \frac{A}{Q}\right)
^{2}\right)  +3+\alpha^{2}\right]  }2\pi r_{h}^{2}.
\end{equation}
Our black hole solution becomes thermodynamically stable /unstable depending
on $C_{Q}>0$ /$C_{Q}<0$ which is not difficult to test from this expression.
For $\left(  \frac{A}{Q}\right)  ^{2}<2$ and $\alpha^{2}<1,$ as an example,
our black hole becomes thermodynamically unstable. Also for $\left(  \frac
{A}{Q}\right)  ^{2}=2$ one gets
\[
C_{Q}=-\frac{\alpha^{2}+1}{3+\alpha^{2}}2\pi r_{h}^{2},
\]
which shows an instability independent of the values of $\alpha.$ Tab. 1
illustrates the stable and unstable regions in terms of $\alpha^{2}$ and
$x=r_{h}^{2}\left(  2-\left(  \frac{A}{Q}\right)  ^{2}\right)  .$
\begin{equation}%
\begin{tabular}
[c]{|l|l|l|l|l|}\hline
& $\alpha^{2}-1<x$ & $-1<x<\alpha^{2}-1$ & $-\left(  3+\alpha^{2}\right)
<x<-1$ & $x<-\left(  3+\alpha^{2}\right)  $\\\hline
$\alpha^{2}<1$ & Unstable & Stable & Stable & Unstable\\\hline
$\alpha^{2}>1$ & Stable & Unstable & Unstable & Stable\\\hline
\end{tabular}
\ \ \tag{Table: 1}\label{Table: 1}%
\end{equation}
Eq. (58) reveals also that $\alpha^{2}=1$ (i.e., the linear dilaton) is a
phase transition point, however, there may be other possible transition points
following a solution for $\alpha$ in the quadratic equation
\begin{equation}
r_{h}^{2}\left(  2-\left(  \frac{A}{Q}\right)  ^{2}\right)  +3+\alpha^{2}=0.
\end{equation}

\section{\bigskip Application of the SCRSM and Hawking Temperature}

In this section, we shall attempt to make a more precise temperature
calculation for the non-extreme LDBHs, $\alpha^{2}=1$ given in Eq. (28), by
using a method of semi-classical radiation spectrum, which has been recently
designated as \textit{SCRSM }\cite{MSH}. The main difference between our
present work with others \cite{CFM, MSH} (and references therein) is that the
considered non-extreme LDBHs possess two horizons, due to having magnetic
charge, instead of one.

Here, we first consider a massless scalar field $\Psi$ with charge $q$ obeying
the covariant Klein-Gordon equation in the LDBH geometry. Namely, we look for
the exact solution of the following equation,%

\begin{equation}
\square\Psi=0,
\end{equation}
where the d' Alembertian operator $\square$ is given by%

\begin{equation}
\square=\frac{1}{\sqrt{-g}}D_{\mu}(\sqrt{-g}g^{\mu\nu}D_{\nu}),
\end{equation}
in which $D_{\mu}$ symbolizes the covariant gauge differential operator as being%

\begin{equation}
D_{\mu}=\partial_{\mu}-iqA_{\mu}.
\end{equation}
The scalar wave function $\Psi$ of Eq. (60) can be separated to the angular
and radial equations by letting%

\begin{equation}
\Psi=Z(r)\pounds (\theta)e^{i(m\varphi-\omega t)},
\end{equation}
the separated angular equation can be found as%

\begin{equation}
\pounds ^{\prime\prime}+\cot\theta\pounds ^{\prime}+\left[  \lambdabar
-\frac{\left(  m+p\cos\theta\right)  ^{2}}{\sin^{2}\theta}\right]  \pounds =0,
\end{equation}
where $p=qQ$ and $\lambdabar$\ is a separation constant. (From now on, a prime
denotes the derivative with respect to its argument.) After setting the
eigenvalue $\lambdabar=l(l+1)-p^{2}$ in Eq. (64), one can see that solutions
to the angular part, $\pounds (\theta)e^{im\varphi},$ are the spin-weighted
spheroidal harmonics $_{p}Y_{lm}(\theta,\varphi)$ with spin-weight $p$
\cite{Gold}.

On the other hand, before proceeding to the radial equation, one may rewrite
the metric function $f(r)$ in Eq. (28) as \ %

\begin{equation}
f(r)=\frac{b}{r}(r-r_{2})(r-r_{1}),
\end{equation}
where $r_{2}$ and $r_{1}$ denote the outer and inner horizons of the LDBHs,
respectively. In the new form of the metric function Eq. (65), the physical
parameters are%

\begin{align}
b  &  =\frac{1}{A^{2}}-2\tilde{V},\text{ }\\
r_{2}  &  =\frac{1}{2b}\left(  c+\sqrt{c^{2}-4ab}\right)  ,\nonumber\\
r_{1}  &  =\frac{1}{2b}\left(  c-\sqrt{c^{2}-4ab}\right)  ,\nonumber
\end{align}
in which%

\begin{equation}
c=\tilde{M}=4M\text{ \ and \ }a=\frac{\lambda_{1}}{\lambda_{2}A^{2}}.
\end{equation}
Since the algorithm in the calculations of the \textit{SCRSM} cover only the
outer region of the black hole $\left(  r>r_{2}\right)  $, we must impose a
condition in order to keep $f(r)$ positive i.e. $b>0$. Henceforth, one can
derive the following radial equation as%

\begin{equation}
b(r-r_{2})(r-r_{1})Z^{\prime\prime}+b(2r-r_{2}-r_{1})Z^{\prime}+(\left(
\frac{r^{2}\omega^{2}}{b(r-r_{2})(r-r_{1})}-\frac{\lambdabar}{A^{2}}\right)
Z=0.
\end{equation}
The above equation can be solved in terms of hypergeometric functions. Here,
we give the final result as%

\begin{align}
Z(r)  &  =C_{1}(r-r_{2})^{i\tilde{\omega}r_{2}}(r-r_{1})^{-i\tilde{\omega
}r_{1}}F\left[  \hat{a},\hat{b};\hat{c};\frac{r_{2}-r}{r_{2}-r_{1}}\right]
+\nonumber\\
&  C_{2}(r-r_{2})^{-i\tilde{\omega}r_{2}}(r-r_{1})^{-i\tilde{\omega}r_{1}%
}F\left[  \hat{a}-\hat{c}+1,\hat{b}-\hat{c}+1;2-\hat{c};\frac{r_{2}-r}%
{r_{2}-r_{1}}\right]  .
\end{align}
The parameters of the hypergeometric functions are%

\begin{equation}
\hat{a}=\frac{1}{2}+i(\frac{\omega}{b}+\sigma),\text{ \ }\hat{b}=\frac{1}%
{2}+i(\frac{\omega}{b}-\sigma),\text{ and\ \ }\hat{c}=1+2i\tilde{\omega}r_{2},
\end{equation}
where%

\begin{equation}
\sigma=\frac{1}{b}\sqrt{\omega^{2}-\frac{\lambdabar b}{A^{2}}-\left(  \frac
{b}{2}\right)  ^{2}},\text{ \ \ }\tilde{\omega}=\omega\tilde{\eta},\text{
\ \ and \ }\tilde{\eta}=\frac{1}{b(r_{2}-r_{1})}.
\end{equation}
Here, $\sigma$\ is assumed to have real values. Furthermore, setting%

\begin{equation}
r-r_{2}=\exp(\frac{x}{\tilde{\eta}r_{2}}),
\end{equation}
one gets the behavior of the partial wave near the outer horizon
($r\rightarrow r_{2}$) as%

\begin{equation}
\Psi\simeq C_{1}e^{i\omega(x-t)}+C_{2}e^{-i\omega(x-t)}.
\end{equation}
One may infer the constants $C_{1}$ and $C_{2}$ as being the amplitudes of the
near-horizon outgoing and ingoing waves, respectively.

In the literature, there exists a useful feature of the hypergeometric
functions, which is a transformation of the hypergeometric functions of any
argument (say $z$) to the hypergeometric functions of its inverse argument
($1/z$). The relevant transformation is given by \cite{AS}%

\begin{align}
F(\bar{a},\bar{b};\bar{c};z)  &  =\frac{\Gamma(\bar{c})\Gamma(\bar{b}-\bar
{a})}{\Gamma(\bar{b})\Gamma(\bar{c}-\bar{a})}(-z)^{-\bar{a}}F(\bar{a},\bar
{a}+1-\bar{c};\bar{a}+1-\bar{b};1/z)\\
&  +\frac{\Gamma(\bar{c})\Gamma(\bar{a}-\bar{b})}{\Gamma(\bar{a})\Gamma
(\bar{c}-\bar{b})}(-z)^{-\bar{b}}F(\bar{b},\bar{b}+1-\bar{c};\bar{b}+1-\bar
{a};1/z).\nonumber
\end{align}
The above transformation leads us to obtain the asymptotic behavior of the
partial wave, easily. After applying the transformation to the general
solution (69), we obtain the partial wave near-infinity as follows%

\begin{equation}
\Psi\simeq\frac{(r-r_{1})^{-i\tilde{\omega}r_{1}}}{\sqrt{r-r_{2}}}\left\{
B_{1}\exp i\left[  \frac{x}{\tilde{\eta}r_{2}}(\sigma+\omega\tilde{\eta}%
r_{1})-\omega t\right]  +B_{2}\exp i\left[  \frac{x}{\tilde{\eta}r_{2}%
}(-\sigma+\omega\tilde{\eta}r_{1})-\omega t\right]  \right\}  .
\end{equation}
On the other hand, since we consider the case of $r\rightarrow\infty,$ the
overall-factor term%

\begin{equation}
(r-r_{1})^{-i\tilde{\omega}r_{1}}\cong\exp i(-\frac{x\omega r_{1}}{r_{2}}),
\end{equation}
whence the partial wave (75) reduces to%

\begin{equation}
\Psi\simeq\frac{1}{\sqrt{r-r_{2}}}\left\{  B_{1}\exp i\left[  \frac{x}%
{\tilde{\eta}r_{2}}\sigma-\omega t\right]  +B_{2}\exp i\left[  -\frac
{x}{\tilde{\eta}r_{2}}\sigma-\omega t\right]  \right\}  ,
\end{equation}
where $B_{1}$ and $B_{2}$ correspond to the amplitudes of the asymptotic
outgoing and ingoing waves, respectively. One can derive the relations between
$B_{1}$,$B_{2}$ and $C_{1}$, $C_{2}$ as follows%

\begin{equation}
B_{1}=C_{1}\frac{\Gamma(\widehat{c})\Gamma(\widehat{a}-\widehat{b})}%
{\Gamma(\widehat{a})\Gamma(\widehat{c}-\widehat{b})}+C_{2}\frac{\Gamma
(2-\widehat{c})\Gamma(\widehat{a}-\widehat{b})}{\Gamma(\widehat{a}-\widehat
{c}+1)\Gamma(1-\widehat{b})},
\end{equation}

\[
B_{2}=C_{1}\frac{\Gamma(\widehat{c})\Gamma(\widehat{b}-\widehat{a})}%
{\Gamma(\widehat{b})\Gamma(\widehat{c}-\widehat{a})}+C_{2}\frac{\Gamma
(2-\widehat{c})\Gamma(\widehat{b}-\widehat{a})}{\Gamma(\widehat{b}-\widehat
{c}+1)\Gamma(1-\widehat{a})}.
\]

Hawking radiation can be considered as the inverse process of scattering by
the black hole such that the outgoing mode at the spatial infinity should be
absent \cite{CFM}. Briefly $B_{1}=0,$ and it naturally yields the coefficient
for reflection by the black hole as%

\begin{equation}
R=\frac{\left\vert C_{1}\right\vert ^{2}}{\left\vert C_{2}\right\vert ^{2}%
}=\frac{\left\vert \Gamma(\widehat{c}-\widehat{b})\right\vert ^{2}\left\vert
\Gamma(\widehat{a})\right\vert ^{2}}{\left\vert \Gamma(1-\widehat
{b})\right\vert ^{2}\left\vert \Gamma(\widehat{a}-\widehat{c}+1)\right\vert
^{2}},
\end{equation}
which is equivalent to%

\begin{equation}
R=\frac{\cosh\pi\left[  \sigma-\frac{\omega}{b}(\frac{r_{2}+r_{1}}{r_{2}%
-r_{1}})\right]  \cosh\pi\left(  \sigma-\frac{\omega}{b}\right)  }{\cosh
\pi\left[  \sigma+\frac{\omega}{b}(\frac{r_{2}+r_{1}}{r_{2}-r_{1}})\right]
\cosh\pi(\sigma+\frac{\omega}{b})}.
\end{equation}
Thus the resulting radiation spectrum is
\begin{equation}
N=\left(  e^{\frac{\omega}{T}}-1\right)  ^{-1}=\frac{R}{1-R}\text{
\ \ }\rightarrow\text{ \ \ }T=\frac{\omega}{\ln(\frac{1}{R})},
\end{equation}
and finally, one can read the more precise value of the temperature as%

\begin{equation}
T=\omega/\ln\left[  \frac{\cosh\pi\left[  \sigma+\frac{\omega}{b}(\frac
{r_{2}+r_{1}}{r_{2}-r_{1}}\right]  \cosh\pi(\sigma+\frac{\omega}{b})}{\cosh
\pi\left[  \sigma-\frac{\omega}{b}(\frac{r_{2}+r_{1}}{r_{2}-r_{1}})\right]
\cosh\pi\left(  \sigma-\frac{\omega}{b}\right)  }\right]  .
\end{equation}
This must be considered as the equilibrium temperature of the quantum field at
the vacuum state valid for all frequencies. In the limit of ultrahigh
frequencies ($\sigma\simeq\frac{\omega}{b}),$ Eq. (82) reduces to
\begin{equation}
T_{high}\simeq\underset{\omega\gg1}{\lim}T\simeq\frac{\omega}{\ln\left[
\exp(\frac{4\pi\omega}{b})\right]  }\simeq\frac{b}{4\pi},
\end{equation}
which smears out the $\omega-$dependence and results in a pure thermal
spectrum. One can immediately observe that Eq. (82) is independent from the
horizons of the non-extreme LDBHs similar to the other $4$-dimensional LDBH
solutions \cite{CFM, MSH} possessing one horizon. But, contrary to the others
\cite{CFM, MSH}, the resulting high frequency temperature $T_{high}$ Eq. (83)
differs from the standard Hawking temperature $T_{H}$ \cite{GR}, which is
computed as usual by dividing the surface gravity by $2\pi:$%

\begin{equation}
T_{H}=\frac{\kappa}{2\pi}=\left.  \frac{f^{\prime}}{4\pi}\right\vert
_{r=r_{2}}=\frac{b}{4\pi}(1-\frac{r_{_{1}}}{r_{_{2}}}).
\end{equation}
Let us note that this same result for the $T_{H}$ can be obtained from
Hawking's period argument of the Euclideanized line element. For this purpose
we complexify time in (29) by $t\rightarrow i\tau$ and rearrange the terms so
that the line element reads in the form $R^{2}\times S^{2},$ given by
\begin{equation}
ds^{2}\sim\left(  \frac{d\rho}{\Sigma_{\circ}}\right)  ^{2}+\left(  \rho
d\tau\right)  ^{2}+\tilde{r}_{\circ h}^{2}\left(  d\theta^{2}+\sin^{2}\theta
d\phi^{2}\right)  .
\end{equation}
Here $\tilde{r}_{\circ h}^{2}$ stands for the value of the radial coordinate
on the outer horizon (when it exists) and the constant $\Sigma_{\circ}$ reads%
\begin{equation}
\Sigma_{\circ}=\frac{1}{2}\left(  \frac{1}{A^{2}}-2\tilde{V}\right)
\left\vert 1-\frac{4\lambda_{1}}{\lambda_{2}A^{2}}\frac{\left(  \frac{1}%
{A^{2}}-2\tilde{V}\right)  }{\sqrt{\tilde{M}+\tilde{M}^{2}-\frac{4\lambda_{1}%
}{\lambda_{2}A^{2}}\left(  \frac{1}{A^{2}}-2\tilde{V}\right)  }}\right\vert
\end{equation}
which relates to the period of the angle $\tau,$ upon the overall
multiplication by $\Sigma_{\circ}$. Since $T_{H}$ is the inverse of the period
we obtain%
\begin{equation}
T_{H}=\frac{1}{2\pi}\Sigma_{\circ}%
\end{equation}
which is identical with (84), valid for double-horizon Hawking temperature. In
order to find the vacuum 'in' and 'out' states for the scalar field we have to
choose the metric such that the surface gravity and mass of the black hole
both vanish. This can be done from (66), by choosing $c^{2}=4ab$ first, to
make an extremal LDBH (with zero temperature), and next, to let $c\rightarrow
0$ to make the mass also zero. These conditions cast our LDBH metric into%
\begin{equation}
ds^{2}=-brdt^{2}+\frac{dr^{2}}{br}+A^{2}rd\Omega^{2}.
\end{equation}
By simple arrangement this vacuum metric transforms into
\begin{equation}
ds^{2}=\rho^{2}\left(  -d\tau^{2}+dx^{2}+d\Omega^{2}\right)
\end{equation}
where%
\[
r=e^{\beta x},\text{ \ \ }t=\frac{\beta}{b}\tau,\text{ \ \ }\rho
=Ae^{\frac{\beta}{2}x},\text{ \ \ }\beta=A\sqrt{b}.
\]
The massless Klein-Gordon equation $\nabla^{2}\Phi=0,$ with $\Phi=\frac
{1}{\rho}\Psi$ takes the form
\begin{equation}
\frac{1}{\rho^{3}}\left(  \partial_{\tau\tau}-\partial_{xx}+\frac{\beta^{2}%
}{4}+\ell\left(  \ell+1\right)  \right)  \Psi=0.
\end{equation}
The vacuum 'in' and 'out' solutions for the scalar field are
\begin{align}
\Phi_{in} &  \sim\frac{1}{\sqrt{r}}e^{-i\left(  \beta\sigma x+\omega t\right)
}\\
\Phi_{out} &  \sim\frac{1}{\sqrt{r}}e^{i\left(  \beta\sigma x-\omega t\right)
}\nonumber
\end{align}
where $\sigma$ has the meaning from (71). Once these states propagate from
vacuum they turn into thermal states as described above.

So the question arises here as in which case does the temperature Eq. (82)
matches with the value of $T_{H}$ in Eq. (84)? The answer is absolutely
related with the value of physical parameter, $\sigma$. Let us assume that the
value of the parameter $\sigma$ is so great that it predominates term
$\frac{\omega}{b}(\frac{r_{2}+r_{1}}{r_{2}-r_{1}})$ (but $\frac{\omega}%
{b}(\frac{r_{2}+r_{1}}{r_{2}-r_{1}})$ is still comparable with $\sigma$) in
the expression of the temperature (82). Unless this assumption is not
violated, the corresponding limit of $T$ will be $T_{H}$. In summary,%

\begin{align}
T_{H}  &  \simeq\underset{\left[  \sigma>\frac{\omega}{b}(\frac{r_{2}+r_{1}%
}{r_{2}-r_{1}})\right]  \gg1}{\lim}T\simeq\omega/\ln\left\{  \frac{\exp
\pi\left[  \sigma+\frac{\omega}{b}(\frac{r_{2}+r_{1}}{r_{2}-r_{1}}\right]
\exp\pi(\sigma+\frac{\omega}{b})}{\exp\pi\left[  \sigma-\frac{\omega}{b}%
(\frac{r_{2}+r_{1}}{r_{2}-r_{1}})\right]  \exp\pi\left(  \sigma-\frac{\omega
}{b}\right)  }\right\} \\
&  \simeq\frac{\omega}{\ln\left\{  \exp2\pi\left[  \frac{\omega}{b}\left(
\frac{r_{2}+r_{1}}{r_{2}-r_{1}}\right)  \right]  \exp2\pi\left(  \frac{\omega
}{b}\right)  \right\}  }\simeq\frac{b}{4\pi}(1-\frac{r_{_{1}}}{r_{_{2}}%
}).\nonumber
\end{align}

Another question may immediately come out: how does $\sigma$ maintain its
predomination against $\frac{\omega}{b}(\frac{r_{2}+r_{1}}{r_{2}-r_{1}})$? To
clarify the question, one can check Eq. (71) in order to see that the
predomination of $\sigma$ strictly\ depends on negative values of
$\lambdabar.$ However, this is possible only with the case of $p^{2}%
=q^{2}Q^{2}>l(l+1)$. Hence, a significant remark is revealed that obtaining
$T_{H}$ of the non-extreme LDBHs from the \textit{SCRSM}, the only possibility
is to consider charged scalar waves instead of chargeless ones.

Furthermore, we want to serve most intriguing figures about the spectrum
temperature Eq. (82). To this end, first we plot $T$ versus frequency $\omega$
of non-extreme LDBHs with $r_{1},r_{2}\neq0$ for low and high $\left\vert
p\right\vert $-values, and display all graphs in Fig. 1. As it can be seen
from Fig. 1 in the high frequencies the thermal behaviors of the LDBHs with
different $\left\vert p\right\vert $-values exhibits similar behaviors in
which their temperatures approach to $T_{high}$ while $\omega\rightarrow
\infty$. The plot with low $\left\vert p\right\vert $-value in Fig. 1 does not
behave like the Hawking temperature. On the other hand, the other plot in Fig.
1, which has high $\left\vert p\right\vert $-value represents the Hawking
temperature $T_{H}$ in the low frequencies ($\omega>0$). Beside this, once the
parameter $\sigma$ is lost its predomination against $\frac{\omega}{b}%
(\frac{r_{2}+r_{1}}{r_{2}-r_{1}})$, the latter plot increases to reach the
$T_{high}$ with increasing frequency as well. In the case of the non-extreme
LDBHs with $r_{1}=0$, there is no difference between $T_{high}$ and $T_{H}$
because of Eq. (84), and at the low $\left\vert p\right\vert $-values the
temperature $T$ exhibits similar behavior as in the case $r_{1},r_{2}\neq0,$
which is the well-known thermal character in the EMD theory \cite{CFM}. By the
way, one should exclude $\omega=0$ during the plotting of the temperature.
Because it causes uncertainty for the temperature Eq. (82) and physically this
case is not acceptable since we consider the propagation of scalar waves. Fig.
2 is about the graph of $T$ versus frequency $\omega$ of non-extreme LDBHs
with $r_{1}=0$ in a high $\left\vert p\right\vert $-value. In this figure, it
is illustrated that by increasing the frequency from $0^{+}$, the temperature
first starts from a constant value, which is $T_{H}$ and then makes a peak
(not much higher than $T_{H}$), and then decreases back to $T_{H}$ while
$\omega\rightarrow\infty$. Rousingly, one can observe that the behavior of the
graph in Fig. 2 is very similar to the graph obtained from the well-known
Planck radiation formula, see for instance \cite{SMM}. Besides, both Fig. 1
and Fig. 2 show us that whenever high $\left\vert p\right\vert $-values are
present, the frequency of the scalar wave needed to detect the temperature of
the LDBHs as to be the Hawking temperature $T_{H}$ can either be very high
(only for $r_{1}=0$ case, which is already known before \cite{CFM}) or low.
The latter information about the relationship between $T_{H}$ and low
frequencies is completely new for us, and may play crucial role for the
thermal detection of the LDBHs in the future.

\section{\bigskip Singularity Analysis}

In section II, we present a solution in $4-$dimensional static spherically
symmetric EMD theory that incorporates two Liouville-type potential terms
coupled with gravity together with magnetically charged dilatonic parameters.
We have clarified that the solution possess a central singularity \ which is a
characteristic feature for spherically symmetric systems. In the solutions
that admit black holes this singularity is clothed by horizons. However, there
are cases that this singularity is not hidden behind a horizons. In such cases
the singularity is called a \textit{naked} singularity.

In classical general relativity, singularities are described as incomplete
geodesics. This simply means that the evolution of timelike or null geodesics
is not defined after a finite proper time. There is a general consensus that a
removal of classical singularities is not important only for quantum gravity
but also for other fundamental theories. In view of this consensus, we are
aiming to analyze whether these classical naked singularities that occur in
the general solution described in Eqs. (16)-(19) and in its linear dilaton
limit given in Eqs. (27) and (28), turn out to be "strong" or "smoothed out"
when probed with quantum test particles. Our analysis will be based on the
pioneering work of Wald \cite{WALD} which was developed by Horowitz and Marolf
\ (HM)\cite{HOROWITZ}. HM, have proposed a criterion to test the classical
singularities with quantum test particles that obey the Klein-Gordon equation
for static spacetimes having timelike singularities. The criterion of HM has
been applied successfully for several spacetimes \cite{HK,PL1,PL2} within the
context of quantum mechanical concepts. Among the others, HM have already
analyzed the quantum singularity for the extreme case of the charged dilatonic
black hole in the absence of Liouville-type potentials. They confirmed that
for a specific interval of dilaton parameter, the singularity is quantum
mechanically regular. The brief review of the criterion is as follows.

A scalar quantum particle with mass $m$ is described by the Klein-Gordon
equation $\left(  \nabla^{\mu}\nabla_{\mu}-m^{2}\right)  \psi=0.$ This
equation can be written by splitting the temporal and spatial portion as
$\frac{\partial^{2}\psi}{\partial t^{2}}=-\mathcal{A}\psi,$ such that the
spatial operator $\mathcal{A}$ is defined by $\mathcal{A}=-\sqrt{f}%
D^{i}\left(  \sqrt{f}D_{i}\right)  +fm^{2},$ where $f=-\xi^{\mu}\xi_{\mu}$
with $\xi^{\mu}$ the timelike Killing field, while$\ D_{i}$ is the spatial
covariant derivative defined on the static slice $\Sigma.$ Then, the
Klein-Gordon equation for a free relativistic particle satisfies
$i\frac{\partial\psi}{\partial t}=\sqrt{\mathcal{A}_{E}}\psi,$ with the
solution $\psi\left(  t\right)  =\exp\left(  it\sqrt{\mathcal{A}_{E}}\right)
\psi\left(  0\right)  .$\ If the extension of the operator $\mathcal{A}$ is
not essentially self-adjoint, the future time evolution of the wave function
is ambiguous. Then, HM criterion defines the spacetime quantum mechanically
singular. However, if there is only one self-adjoint extension, the operator
$\mathcal{A}$ is said to be\ essentially self-adjoint and the quantum
evolution $\psi\left(  t\right)  $ is uniquely determined by the initial
condition. According to the HM criterion, this spacetime is said to be quantum
mechanically regular. Consequently, a sufficient condition for the operator
$\mathcal{A}$ to be essentially self-adjoint is to investigate the solutions
satisfying the following equation ( see Ref. \cite{IH} for a detailed
mathematical background),
\begin{equation}
\mathcal{A}\psi\pm i\psi=0.
\end{equation}
This equation admits separable solution and hence the radial part becomes,%

\begin{equation}
\frac{\partial^{2}\phi}{\partial r^{2}}+\frac{1}{fR^{^{2}}}\frac
{\partial\left(  fR^{2}\right)  }{\partial r}\frac{\partial\phi}{\partial
r}-\frac{l\left(  l+1\right)  }{fr^{2}}\phi-\frac{m^{2}}{f}\phi\pm i\frac
{\phi}{f^{2}}=0,
\end{equation}
in which $l\left(  l+1\right)  \geq0$ is the eigenvalue of the Laplacian on
the $2-$sphere. The necessary condition for the operator $\mathcal{A}$ to be
essentially self adjoint is that at least one of the solutions to this
equation fails to be of finite norm when $r\rightarrow0.$ In summery, the self
adjointness of the operator $\mathcal{A}$, implies the well-posedness of the
initial value problem. Therefore, the suitable norm $\left\Vert \phi
\right\Vert $ for this case is the Sobolev norm which is used first time
within this context by Ishibashi and Hosoya \cite{IH} defined by,%

\begin{equation}
\left\Vert \phi\right\Vert ^{2}=\frac{q^{2}}{2}\int R^{2}f^{-1}\left\vert
\phi\right\vert ^{2}d\mu dr+\frac{1}{2}\int R^{2}f\mid\left\vert
\frac{\partial\phi}{\partial r}\right\vert ^{2}d\mu dr,
\end{equation}
where $q^{2}$ is a positive constant and $d\mu$ is the volume element on the
unit $2-$sphere. The regularity of the central singularity at $r=0$ in quantum
mechanical sense requires that the squared norm of the solutions of the Eq.
(94) should be divergent for each $l\left(  l+1\right)  $ and each sign of
imaginary term. The norm $\left\Vert \phi\right\Vert $ is divergent for
$l\left(  l+1\right)  >0$ if it is for $l=0$, so essential self-adjointness
will be examined for $l=0$ ($S-wave$) case. This implies essential self
adjointness for the operator $\mathcal{A}$. Furthermore, we assume, a massless
case (i.e. $m=0$), and ignoring the term $\pm i\frac{\phi}{f^{2}}$ ( since it
is negligible near the origin).

\subsection{A more general case:}

The general solution for any value of $\alpha^{2}$ which is related to the
dilaton parameters $\gamma_{1}$ and $\gamma_{2}$ is given in the Eq. (18).
Since this solution is complicated enough for integrability, we consider the
specific values of $\alpha^{2}=3$ and $V_{1}=0.$ Hence, the general solution becomes;%

\begin{equation}
f(r)=\frac{16a_{2}}{\sqrt{r}}\left(  r-r_{2}\right)  \left(  r-r_{1}\right)  ,
\end{equation}
where%

\begin{align}
\tilde{r}_{1,2}  &  =\frac{M\pm\sqrt{M^{2}-4a_{1}a_{2}}}{2a_{2}},\text{
\ \ \ \ \ }\\
a_{1}  &  =\frac{Q^{2}\lambda_{1}}{4A^{4}},\text{ \ \ \ \ \ \ \ \ \ \ }%
a_{2}=\frac{1}{12}\left(  \frac{Q^{2}\lambda_{2}}{A^{4}}-V_{2}\right)
.\nonumber
\end{align}
The extreme case occurs when $M^{2}=4a_{1}a_{2}.$ In this case there is one
horizon only and it is given by $r_{h}=\frac{M}{2a_{2}}.$ If $M>\sqrt
{M^{2}-4a_{1}a_{2}}$ and $M^{2}>4a_{1}a_{2},$ this particular case admits two
horizons given by $\tilde{r}_{1,2}.$ However, if $M^{2}-4a_{1}a_{2}<0,$ no
black hole forms and hence, the singularity at $r=0$ becomes naked.

As a requirement of the HM criterion, the singularity at $r=0$ must have a
timelike character. This can be checked if one introduces tortoise coordinate
defined by $r_{\ast}=\int\frac{dr}{f}$ and take its limit as $r\rightarrow0.$
We found that the limit is finite. Therefore, the singularity is timelike. The
solution for Eq. (94) is%

\begin{equation}
\phi(r)=\frac{A^{-2}}{16a_{2}\left(  \tilde{r}_{2}-\tilde{r}_{1}\right)  }%
\ln\left\vert \frac{r-\tilde{r}_{2}}{r-\tilde{r}_{1}}\right\vert .
\end{equation}
The first and the second terms of the squared norm (95) is finite, when
$r\rightarrow0.$

Consequently, the operator $\mathcal{A}$ is not essentially self-adjoint and
therefore, the central singularity $r=0,$ remain quantum mechanically singular.

\subsection{The linear dilaton case:}

The metric function for the linear dilaton case can be written as, (from Eq. (65))%

\[
f(r)=\frac{b}{r}\left(  r-r_{2}\right)  \left(  r-r_{1}\right)  ,
\]
where
\begin{align*}
r_{1,2} &  =\frac{\tilde{M}\pm\sqrt{\tilde{M}^{2}-4ab}}{2b},\\
a &  =\frac{\lambda_{1}}{A^{2}\lambda_{2}}\text{ ,\ \ \ \ \ \ \ \ \ \ \ }%
b=\left(  \frac{1}{A^{2}}-2\tilde{V}\right)  .
\end{align*}
The naked singularity occurs when $\tilde{M}^{2}-4ab<0.$ The tortoise
coordinate $r_{\ast}=\int\frac{dr}{f}$ is finite and indicating a timelike
character at $r=0$. The Penrose diagram of this particular case is shown in
Fig. 3-a. The radial part of the separable Eq. (94) has solution for the
linear dilaton case as,%

\begin{equation}
\phi(r)=\frac{1}{bA^{2}\left(  r_{2}-r_{1}\right)  }\ln\left\vert
\frac{r-r_{2}}{r-r_{1}}\right\vert .
\end{equation}

The first and the second term of the squared norm defined in Eq. (95) is
finite. Therefore the spacetime is quantum mechanically singular. For the
double-horizon case, $\tilde{M}^{2}-4ab>0,$ which implies $r_{1}\neq r_{2}%
\neq0\neq r_{1},$ the timelike singularity at $r=0$ is not naked, and its
Penrose diagram is depicted in Fig. 3-b.\ However, for a special case
$\lambda_{1}=0,$ the solution to Eq. (94) is%

\begin{equation}
\phi(r)=\frac{A^{2}}{b\tilde{M}}\ln\left\vert \frac{r-r_{h}}{r}\right\vert ,
\end{equation}
in which $r_{h}=\frac{\tilde{M}}{b}.$ The first term of the squared norm (95)
is finite, whereas the second term behaves as,%

\begin{equation}
\sim\left.  (\ln\left\vert r\right\vert )\right\vert _{r=0}\rightarrow\infty.
\end{equation}
Hence, under the condition $\lambda_{1}=0,$ the central classical singularity
becomes quantum mechanically non-singular. When we have a single-horizon, with
the choice $\lambda_{1}=0$, for example, the singularity $r=0$, is shown in
the Penrose diagram (Fig. 3-c).

\subsection{Near horizon behaviors}

In order to study the global behavior of our solution, at least for specific
choices of parameters, and to be able to sketch the Penrose diagrams, we cast
the metric into the form apt for near horizons. With the choice $V_{1}%
=V_{2}=0$ our metric function $f(r)$ takes the form%
\begin{equation}
f\left(  r\right)  =\frac{1}{A^{2}r^{1+\delta}}\left(  r^{2}-\frac{4M_{QL}%
}{1+\delta}r+\frac{\lambda_{1}}{\lambda_{2}}\frac{1-\delta}{1+\delta}\right)
,
\end{equation}
in which
\begin{equation}
\delta=\frac{1-\alpha^{2}}{1+\alpha^{2}},\text{ \ }-1<\delta<1
\end{equation}
and
\begin{equation}
A^{2}=\frac{2}{1-\delta}\lambda_{2}Q^{2}.
\end{equation}
Upon the choice of parameters involved, we can have, double, single or
no-horizon cases. By a redefinition for time, our line element reads, in
brief,%
\begin{equation}
ds^{2}=A^{2}d\tilde{s}^{2}%
\end{equation}
in which
\begin{equation}
d\tilde{s}^{2}=-\frac{\left(  r-r_{-}\right)  \left(  r-r_{+}\right)
}{r^{1+\delta}}dt^{2}+\frac{r^{1+\delta}}{\left(  r-r_{-}\right)  \left(
r-r_{+}\right)  }dr^{2}+r^{1+\delta}d\Omega^{2}%
\end{equation}
and
\begin{equation}
r_{\pm}=\frac{2M_{QL}}{1+\delta}\left(  1\pm\sqrt{1-\frac{\lambda_{1}}%
{\lambda_{2}}\frac{1-\delta^{2}}{4M_{QL}^{2}}}\right)  .
\end{equation}
We note that the global structure of $d\tilde{s}^{2}$ is same with $ds^{2},$
and therefore we analyze $d\tilde{s}^{2}.$ The singularity structure of (106)
can be seen from the Kretchmann scalar-K, which reads%
\begin{equation}
\lim_{r\rightarrow0}K\sim\left\{
\begin{array}
[c]{cc}%
r^{-2\left(  \delta+3\right)  }, & \delta\neq\pm1\\
r^{-8}, & \delta=+1
\end{array}
\right.
\end{equation}%
\begin{equation}
\lim_{r\rightarrow\infty}K\sim\left\{
\begin{array}
[c]{cc}%
r^{-2\left(  \delta+1\right)  }, & \delta\neq\pm1\\
\text{constant}, & \delta=-1
\end{array}
\right.
\end{equation}
We concentrate ourselves now to the near horizon geometry by the following
reparametrization, with new coordinates ($\tilde{r},\tilde{t}$)
\begin{equation}
r_{-}=r_{\circ},\text{ \ \ }r_{+}=r_{\circ}+\epsilon b_{0},\text{
\ \ }r=r_{\circ}+\epsilon\tilde{r},\text{ \ \ }t=\frac{1}{\epsilon}\tilde{t}%
\end{equation}
in which $r_{\circ}$ and $b_{0}$ are constants and $\epsilon\rightarrow0,$ is
a small parameter. We obtain, upon relabeling $\tilde{r}=r$ and $\tilde{t}=t$%
\begin{equation}
d\tilde{s}^{2}=-\frac{r\left(  r-b_{0}\right)  }{r_{\circ}^{1+\delta}}%
dt^{2}+\frac{r_{\circ}^{1+\delta}}{r\left(  r-b_{0}\right)  }dr^{2}+r_{\circ
}^{1+\delta}d\Omega^{2}.
\end{equation}
In Fig. 4 we plot the Penrose diagrams for the specific cases $b_{0}=0$, and
$b_{0}>0$. The case $b_{0}<0$ doesn't differ from the case of $b_{0}=0,$ and
as a matter of fact this particular case corresponds to the BR limit, which is
known to correspond to the near extreme geometry of the RN black hole. The
more standard BR is obtained from the present one by the inversion
$r\rightarrow\frac{1}{r}.$

\section{CONCLUSION}

We have shown that dilaton field with Liouville's potential interpolates
between RN black hole and non-black hole BR solution. The general solution for
the metric function suggests that dilatonic presence induces significant
changes in the solutions; for example, asymptotically flat black holes become
non-asymptotically flat . It is shown, through radial linear perturbation,
that dilaton can add instabilities to the otherwise stable RN black hole
whereas BR remains stable. From the thermodynamic point of view also, by
invoking specific heat the system can be tested against stability and phase
transition. In the non-extreme LDBH case, which is a particular solution of
our general solution the statistical and the standard Hawking temperatures are
compared and plotted. It has been pointed out that with charged scalar waves
and spin-weighted coupling the two results match for the case of double
horizons. We recall that in the single horizon case, in spite of the existence
of a linear dilaton such a discrepancy does not arise. It is remarkable that
the spin-weighted spheroidal harmonics serve to convert the diverging
temperature spectrum into a finite one. The presence of dilaton makes the
spacetime highly singular at $r=0$. Whether these singularities are quantum
mechanically singular also, or not, we send a quantum test particle and apply
the criterion due to Horowitz and Marolf. We find that under certain choice of
our parameters the naked singularities create an infinite repulsive quantum
potential so that the particle feels a regular space time.

\bigskip

\textbf{FIGURE CAPTIONS}

\textbf{Figure 1}: Temperature $T$ as a function of $\omega$ for the
non-extreme LDBHs in the case of $r_{1},r_{2}\neq0$. The plots are governed by
Eq. (82). Different line styles belong to different $\left\vert p\right\vert
$-values: Dotted line corresponds to $\left\vert p\right\vert =0.5$ (as an
example of low $\left\vert p\right\vert $-values) and solid line is for
$\left\vert p\right\vert =10$ (as an example for high $\left\vert p\right\vert
$-values). The physical parameters in Eq. (82) are chosen as follows: $l=1$,
$b=1,$ $A=1$, $r_{1}=0.5$ and $r_{2}=1.$

\textbf{Figure 2}: Temperature $T$ as a function of $\omega$ for the
non-extreme LDBHs in the case of $r_{2}\neq0$ and $r_{1}=0,$ and when $p$ has
a high value. The plot is governed by Eq. (82). The physical parameters in Eq.
(82) are chosen as follows: $\left\vert p\right\vert =10$, $l=1$, $b=1,$ $A=1$
and $r_{2}=1.$

\textbf{Figure 3-a}: Penrose diagram for no-horizon case, $\tilde{M}%
^{2}-4ab<0$ in which $r=0$ is a naked singularity.

\textbf{Figure 3-b}: Penrose diagram of the LDBH with two distinct horizons
$r_{1}\neq r_{2}$, where $r=0$ is a timelike singularity.

\textbf{Figure 3-c}: Penrose diagram of the LDBH with a single horizon at
$r=r_{h}.$ singular nature of $r=0$ is not affected.

\textbf{Figure 4-a}: Penrose diagram for the line element (111) with $b_{0}%
=0$. There is no singularity and no horizon. By inverting coordinate
$r\rightarrow\frac{1}{r}$ we obtain the standard BR diagram. We note that the
choice $b_{0}<0$ is also similar to this case.

\textbf{Figure 4-b}: For $b_{0}>0$ there is a horizon and the Penrose diagram
is as shown with singularities at the null boundaries. By inversion as in
(3-a) we interchange $r=0$ $(r\rightarrow\infty)$ with $r\rightarrow\infty$
$(r=0).$

\end{document}